\documentclass[a4paper,11pt]{article}
\usepackage{pos}
\usepackage{amsmath}
\usepackage{comment}
\usepackage{gensymb}
\usepackage{float}
\usepackage{placeins}
\usepackage{url}
\usepackage{acronym}
    \newacro{UL}{upper limit}
    \newacro{VHE}{very high-energy}
\usepackage{lineno}

\title{Recent MAGIC results on Galactic binaries}
    \ShortTitle{Recent MAGIC results on Galactic binaries}

\author*[a]{E. Molina}
\author[b]{A. L\'opez-Oramas}
\author[c]{D. Hadasch}
\author[d,e]{J. Hoang}

\affiliation[a]{Departament de F\'isica Qu\`antica i Astrof\'isica,
                Institut de Ci\`encies del Cosmos (ICCUB),
                Universitat de Barcelona (IEEC-UB), \\
                Mart\'i i Franqu\`es 1, 08028 Barcelona, Spain}
\affiliation[b]{Instituto de Astrof\'isica de Canarias and Dpto. de  Astrof\'isica,
                Universidad de La Laguna, \\
                E-38200, La Laguna, Tenerife, Spain}
\affiliation[c]{Institute for Cosmic Ray Research (ICRR),
                The University of Tokyo, \\
                Kashiwa, 277-8582 Chiba, Japan}
\affiliation[d]{IPARCOS Institute and EMFTEL Department,
                Universidad Complutense de Madrid, \\
                E-28040 Madrid, Spain}
\affiliation[e]{Department of Astronomy,
                University of California Berkeley, \\
                Berkeley CA 94720, USA}

\forColl{MAGIC}

\emailAdd{emolina@fqa.ub.edu}

\newcommand{\maxi}{MAXI~J1820+070 }
\newcommand{\hess}{HESS~J0632+057 }
\newcommand{\bex}{1A~0535+262 }
\newcommand{\swift}{\textit{Swift}/BAT }

\abstract{
X-ray and gamma-ray binaries are systems consisting of a compact object and normally a non-degenerate companion star. Most of these sources have been shown to emit radiation in a broad frequency range, from radio up to X-rays and sometimes gamma rays. We report on recent results in very high-energy gamma rays above 100~GeV obtained by the MAGIC Collaboration for the Galactic X-ray binaries \maxi and \bex\!, and the gamma-ray binary \hess\!. Multiwavelength data at lower energies are also provided for a better contextualisation of the sources.
}

\FullConference{37$^{\rm{th}}$ International Cosmic Ray Conference (ICRC 2021)\\
		July 12th -- 23rd, 2021\\
		Online -- Berlin, Germany}

\begin{document}
\maketitle

\section{Introduction}\label{sec:intro}

There are a number of binary systems in our Galaxy, typically consisting of a compact object and a non-degenerate companion star, that produce X-ray and gamma-ray emission. Depending on the energy at which their emission peaks, they are called either X-ray or gamma-ray binaries. Two main scenarios have been proposed to explain the observed radiation, one involving matter accretion and jet launching by the compact object (microquasar scenario), and another one in which the compact object is a pulsar that interacts with the star through their winds (pulsar wind scenario).

\maxi is a low-mass microquasar with a black hole that was discovered in X-rays on March 11 2018 (MJD~58188.8) by the MAXI instrument \cite{kawamuro18}. Soon after its discovery, \maxi reached an exceptionally high X-ray flux peaking at $\sim 4$ times that of the Crab Nebula in the 15--50~keV range \cite{delsanto18,shidatsu19}. The distance to the source was set to $2.96 \pm 0.33$~kpc \cite{atri20}, and an orbital period of $16.4518 \pm 0.0002$~h was found \cite{torres19}. Regarding its X-ray state, \maxi followed the usual cycle for microquasars with a black hole in outburst \cite{fender16}, going from a hard spectral state (HS) to a soft state (SS), and back to the HS shortly before going into quiescence \cite{shidatsu19}. The bottom panel in Fig.~\ref{fig:maxiLC} allows to visualise the X-ray state evolution of the source throughout the outburst.

\hess is a gamma-ray binary consisting of a compact object and a massive Be-type star. It was discovered in \ac{VHE} gamma-rays as an unidentified point-like source during H.E.S.S. observations of the Monoceros region \cite{hess07_hessj0632}, and since then it has been detected at different frequencies from radio to high-energy gamma-rays. The distance to the source was set to $1.1 - 1.7$~kpc \cite{aragona10}, and the orbital period to $316.8^{+2.6}_{-1.4}$~days from X-ray data \cite{malyshev19}. The pulsar-wind scenario has been proposed for \hess \cite{moritani15,barkov18}, although the microquasar scenario cannot be ruled out.

\bex is a Be X-ray binary composed of a giant Be-type star and a pulsar. This source has undergone periodic X-ray flares every few years since its discovery in 1975 (see \cite{caballero07} and references therein). In November 2020, it displayed an especially bright X-ray outburst reported by \swift \cite{Mandal2020ATel14157....1M} and confirmed by MAXI \cite{MAXI2020ATel14173....1N}. This is the brightest X-ray event recorded from this source up to date, reaching a luminosity level of 12 Crab in the 15--50~keV band in 19 November 2020. \bex was detected in radio for the first time during the 2020 outburst \cite{eijnden20}, indicating the presence of non-thermal emission mechanisms. The distance to the source is $2.1 \pm 0.2$~kpc \cite{treuz18}, and it has an orbital period of $111 \pm 0.4$~days \cite{priedhorsky83}. Super-critical accretion onto the pulsar during the outburst was reported by \cite{Jaisawal2020ATel14227....1J}.

In these proceedings, we provide an overview of the latest results on Galactic binary systems obtained with the MAGIC telescopes at energies above 100 GeV. In particular, we present results on the three sources described above: \maxi\!, \hess\ and \bex\!. We put these observations into a multi-wavelength context and discuss their physical implications for the different observed sources. The structure of the proceedings is the following: Sect.~\ref{sec:obs} describes the observations performed. In Sect.~\ref{sec:results}, the results for each source are shown and a discussion is provided. Finally, a short summary is given in Sect.~\ref{sec:summary}. 

\section{Observations and data analysis with MAGIC}\label{sec:obs}

MAGIC \cite{magic16_hardware} is a stereoscopic system of two Imaging Atmospheric Cherenkov Telescopes located at the Roque de los Muchachos Observatory in La Palma, Spain ($29^\circ$N, $18^\circ$W, 2200~m above sea level). Each telescope has a 17-m diameter mirror dish that collects and focuses light into its focal plane, where a fast photo-multiplier tube camera with a $3.5^\circ$ field of view is located. The analysis of the MAGIC data presented here was done following the standard procedure described in \cite{magic16_performance}.

\subsection{\maxi}
\maxi was observed with the MAGIC telescopes during the X-ray outburst from March to October 2018, covering the initial HS of the source as well as the state transitions. A total of 22.5~h of data survived the quality cuts, all of them taken under dark conditions. These observations were performed together with the H.E.S.S. and VERITAS Collaborations for an added total of 61~h of good-quality data. In these proceedings we only report on the MAGIC observations, for which a summary can be found in Table~\ref{tab:maxiObs}. We refer the interested reader to an upcoming publication for the details of the joint observational campaign of MAGIC, H.E.S.S. and VERITAS.

\begin{table}
    \begin{center}
    \caption{Summary of the observations of \maxi with the MAGIC telescopes, after data quality cuts. The effective observation time and zenith angle range are shown for each source state, as well as for the whole data set.}
    \begin{tabular}{c c c}
    \hline \hline
    Source state	        & Time [h]	& Zenith angle [deg]   \\
    \hline
    Hard State              & 14.2		& 21 -- 58             \\
    HS $\rightarrow$ SS     & 4.9		& 21 -- 48             \\
    SS $\rightarrow$ HS     & 3.4		& 28 -- 56             \\
    \hline
    TOTAL                   & 22.5		& 21 -- 58             \\
    \hline
    \end{tabular}
    \label{tab:maxiObs}
    \end{center}
\end{table}

\subsection{\hess}
The gamma-ray binary \hess was observed with MAGIC between October 2010 and November 2017, accumulating a total of 68~h of data. Some of these data were already published in \cite{magic12_hessj0632}. After quality cuts, a total of 57.4~h remain. The data were taken under dark and moderate-to-strong moonlight conditions, and with zenith angles between 38 and 67$^\circ$. The presence of a bright night sky background for the moon observations increases their energy threshold with respect to dark observations from 147 to 251 GeV. The MAGIC observations of \hess are part of a larger joint campaign with H.E.S.S. and VERITAS.

\subsection{\bex}
\bex was observed with MAGIC between November 17 and December 19 2020, contemporaneously to the giant X-ray flare. The bulk of the MAGIC observations took place during the peak of X-ray emission around November 19 (MJD~59172). After the data quality selection, which accounts for non-optimal weather conditions, 8.2~h of observations remained for further analysis, all of them done under dark conditions and at zenith angles below 35$^\circ$.

\section{Results and discussion}\label{sec:results}

\subsection{\maxi}\label{sec:results_maxij1820}

The MAGIC observations of the outburst of \maxi do not yield a significant detection of the source, and an integral flux \ac{UL} above 200~GeV of $2.2 \times 10^{-12}$~cm$^{-2}$~s$^{-1}$ is obtained for the full data set. This value is computed using a $95\%$ confidence level and assuming a power-law spectrum with index $-2.5$.

\begin{figure}
    \begin{center}
        \includegraphics[width=0.9\textwidth]{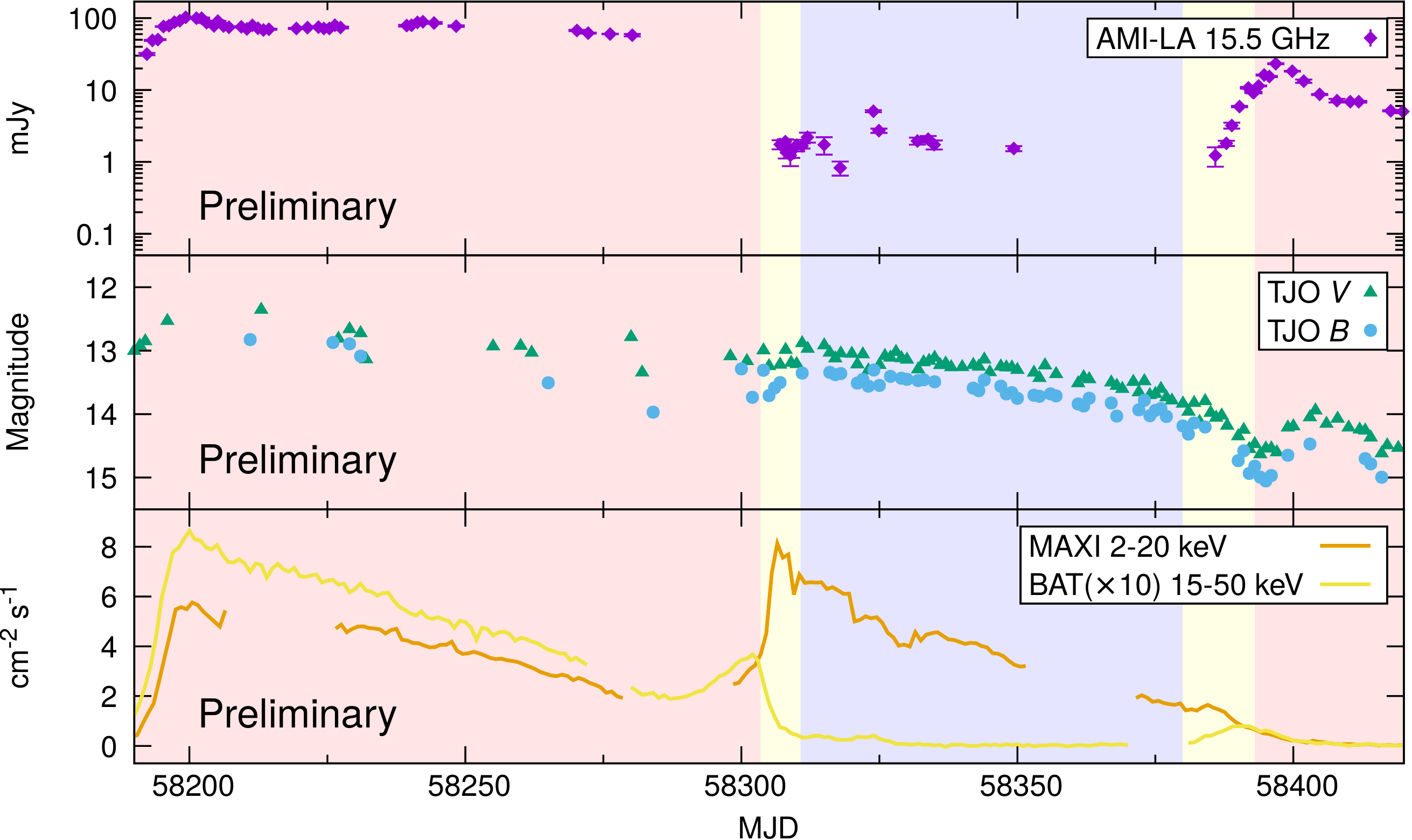}
        \caption{From \textit{top} to \textit{bottom}: Radio, optical and X-ray light curves of \maxi during its 2018 outburst. The shaded areas correspond to the HS (light red), the HS--SS and SS--HS transitions (yellow), and the SS (light blue). The units of the bottom panel are [ph cm$^{-2}$ s$^{-1}$] for MAXI/GSC, and [counts cm$^{-2}$ s$^{-1}$] for \swift\!. The latter fluxes are multiplied by 10 for a better visualisation.}
        \label{fig:maxiLC}
    \end{center}
\end{figure}

The light curve of \maxi during the outburst, as seen by different instruments, is shown in Fig.\ref{fig:maxiLC}. Radio data from AMI-LA is taken from \cite{bright20}, while optical data in the $V$ and $B$ filters from the Joan Or\'o Telescope (TJO; \cite{colome10}) are taken from \cite{celma19}. We also include public light curves from MAXI/GSC\footnote{\url{http://maxi.riken.jp/star_data/J1820+071/J1820+071.html}} and \swift\!\footnote{\url{https://swift.gsfc.nasa.gov/results/transients/weak/MAXIJ1820p070}}. Without accounting for the radio emission of transient ejections launched during the HS--SS transition, which are dominant throughout the SS \cite{bright20}, the radio and hard X-ray fluxes have similar behaviours.
They follow the standard picture of black-hole X-ray binaries, for which steady radio jets in the HS coexist with a hard X-ray emitting corona, both of them disappearing in the SS. These multiwavelength observations, together with the obtained \ac{VHE} \acp{UL}, allow to set constraints on the size and position of a potential gamma-ray emitter in \maxi\!. A detailed discussion on this issue will be given in a future joint publication of MAGIC, H.E.S.S. and VERITAS.

\subsection{\hess}\label{sec:results_hessj0632}

\begin{figure}
    \begin{center}
        \includegraphics[width=0.8\textwidth]{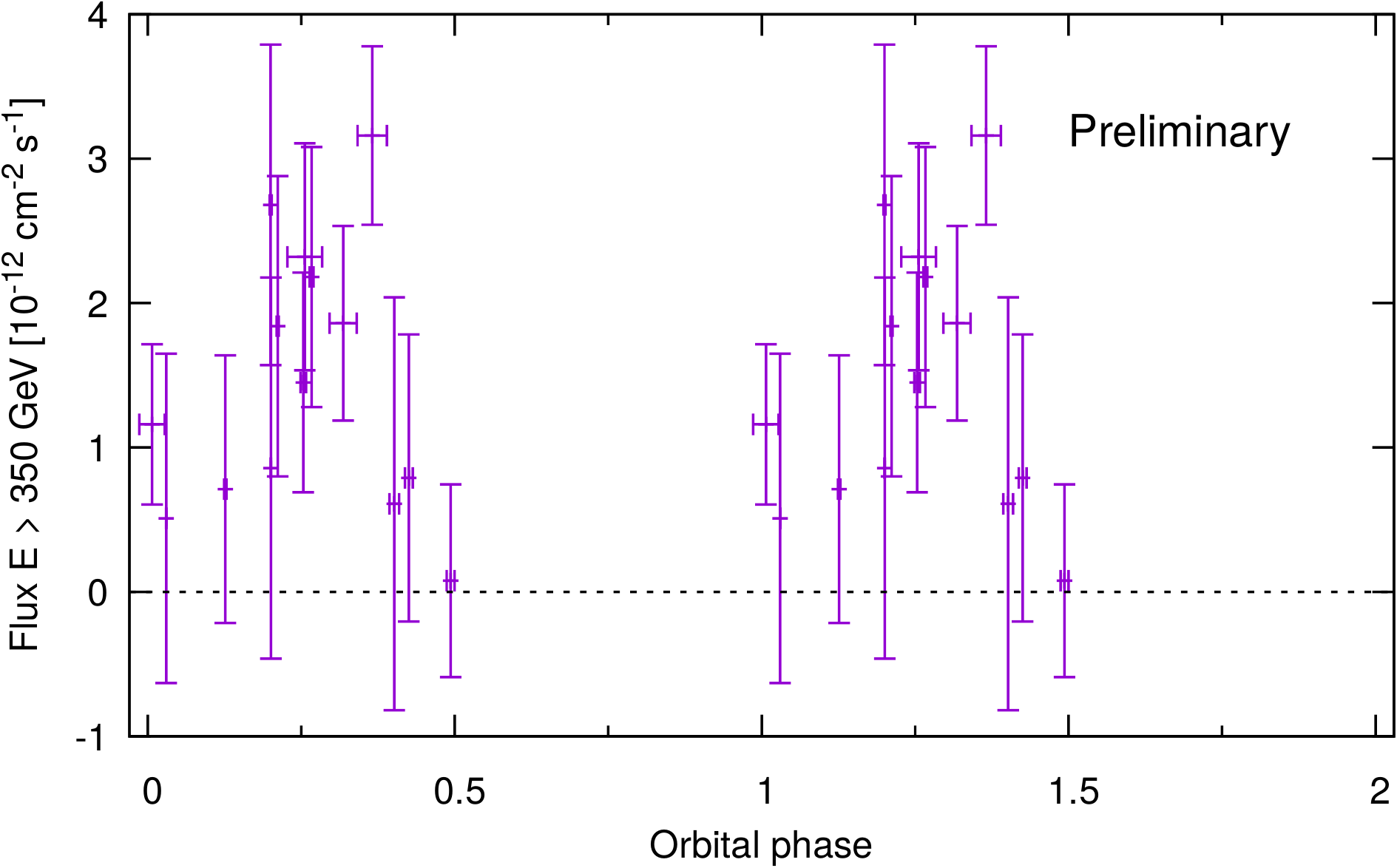}
        \caption{Phase-folded light curve of \hess assuming an orbital period of 316.8~days at energies above 350 GeV, as seen by MAGIC. Two orbits are shown for visualisation purposes.}
        \label{fig:hessjLC}
    \end{center}
\end{figure}

The long-term observations of \hess have allowed to better sample its \ac{VHE} emission along the orbit. The \ac{VHE} light curve obtained from MAGIC data above 350 GeV is shown in Fig.~\ref{fig:hessjLC}, phase folded with an orbital period of 316.8~days. The observations were mainly done at orbital phases between 0 and 0.5. A maximum in the flux is found in the orbital range 0.2--0.4, and a minimum is present at phases around 0.5.

The results of the \hess analysis are part of a larger long-term observational campaign involving also the H.E.S.S. and VERITAS telescopes. This multi-year sampling of \hess allowed for the first-time determination of the orbital period from \ac{VHE} data only, consistent with the previously determined X-ray period of 316.8~days \cite{malyshev19}. A forthcoming publication will show the combined results of the three Collaborations, and will also include X-ray and optical data. We refer to this upcoming joint publication for a detailed discussion of the observations and the interpretation of the results.

\subsection{\bex}\label{sec:results_1a0535}

\begin{figure}
    \begin{center}
        \includegraphics[width=0.9\textwidth]{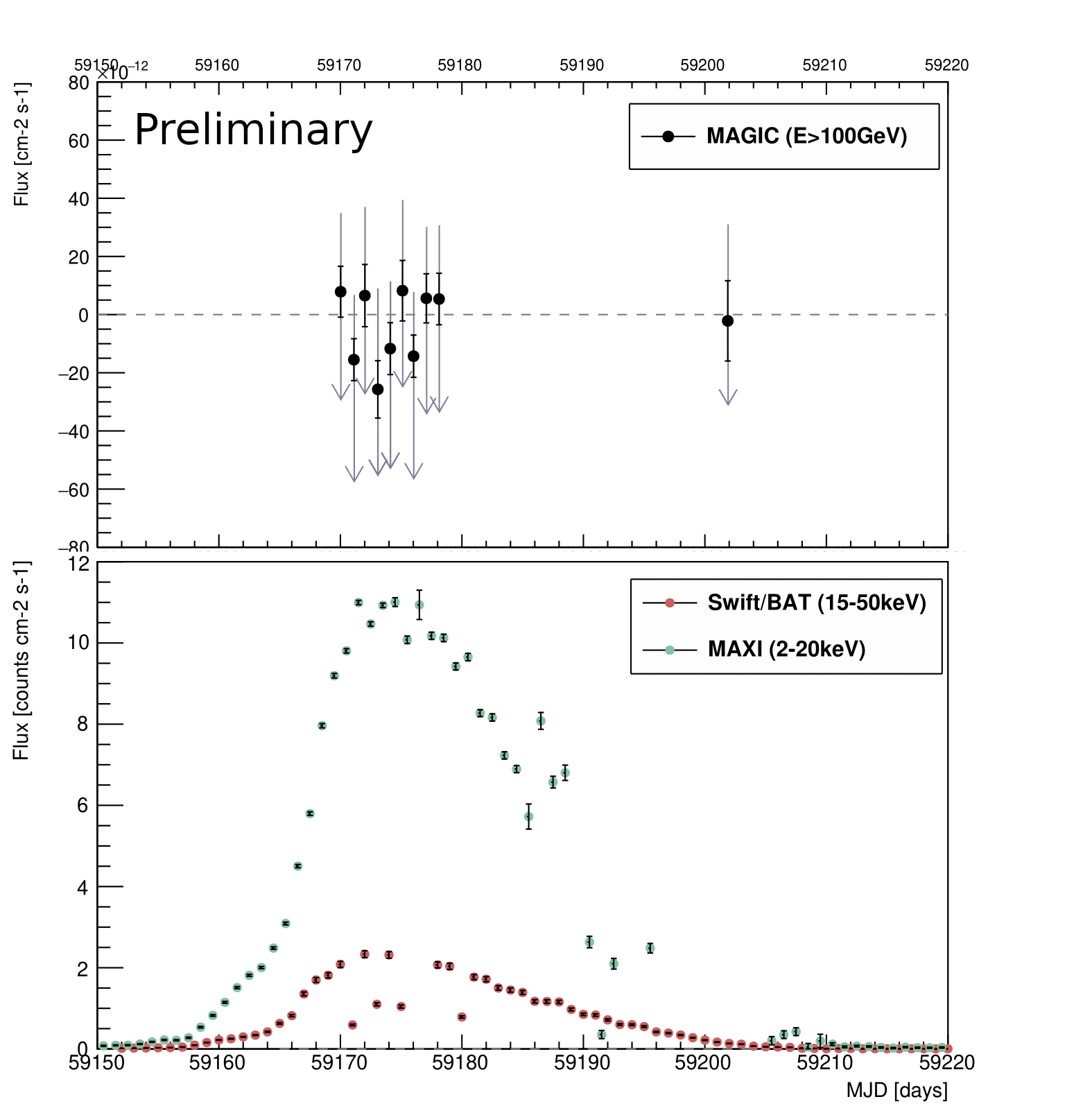}
        \caption{\textit{Top panel}: \ac{VHE} light curve of \bex above 100~GeV for the source outburst in 2020. The points show the computed fluxes, whereas the grey arrows represent the corresponding \acp{UL} at a $95\%$ confidence level. \textit{Bottom panel}: X-ray light curve of \bex as seen by \swift and MAXI in the same time period. Both panels use a daily time binning.}
        \label{fig:1a0535LC}
    \end{center}
\end{figure}

The analysis of the 8.2~h of observations of \bex does not reveal any significant \ac{VHE} signal in the MAGIC data. The detection significance of the whole data set is 1.3$\sigma$ (computed following the usual Li\&Ma formula \cite{lima83}). The integral flux \ac{UL} at energies above 100 GeV is 5.0$\times10^{-12}$~cm$^{-2}$~s$^{-1}$ (with a $95\%$ confidence level and an assumed power-law spectral index of $-2.6$). Fig.~\ref{fig:1a0535LC} shows the light curve of \bex in \ac{VHE} gamma rays and X-rays, the latter extracted from public \swift\!\footnote{\url{https://swift.gsfc.nasa.gov/results/transients/1A0535p262/}} and MAXI\footnote{\url{http://maxi.riken.jp/star_data/J0538+263/J0538+263.html}} data. All the individual \ac{VHE} flux points have a significance below $2.5\sigma$, and the corresponding \acp{UL} are also shown.

\section{Summary}\label{sec:summary}

We have reported on recent \ac{VHE} gamma-ray obtained with the MAGIC telescopes for the Galactic binary sources \maxi\!, \hess and \bex\!. For the first two sources, upcoming joint publications with H.E.S.S. and VERITAS will provide more details on the results and the corresponding discussion. \maxi\! is not detected at \ac{VHE} and flux \acp{UL} are computed, which together with the multiwavelength data from radio to X-rays allow to put constraints on a potential \ac{VHE} emitter in the source. A clear orbital modulation is observed in \hess at \ac{VHE}, and the combined MAGIC, H.E.S.S. and VERITAS observations of the source allowed to determine for the first time its orbital period from \ac{VHE} data only. Finally, the brightest recorded X-ray flare from \bex has not come with detectable \ac{VHE} emission, and integral flux \acp{UL} are reported.

\footnotesize\acknowledgments

We acknowledge the support from the agencies and organisations listed here:

\url{https://magic.mpp.mpg.de/acknowledgments_ICRC2021}

\FloatBarrier
\bibliographystyle{JHEP}
\small \bibliography{bibliography}

\section*{The MAGIC Collaboration}
\scriptsize
\noindent
V.~A.~Acciari$^{1}$,
S.~Ansoldi$^{2,41}$,
L.~A.~Antonelli$^{3}$,
A.~Arbet Engels$^{4}$,
M.~Artero$^{5}$,
K.~Asano$^{6}$,
D.~Baack$^{7}$,
A.~Babi\'c$^{8}$,
A.~Baquero$^{9}$,
U.~Barres de Almeida$^{10}$
% ...
\\

\noindent
$^{1}$ {Instituto de Astrof\'isica de Canarias and Dpto. de  Astrof\'isica, Universidad de La Laguna, E-38200, La Laguna, Tenerife, Spain} 
$^{2}$ {Universit\`a di Udine and INFN Trieste, I-33100 Udine, Italy} 
$^{3}$ {National Institute for Astrophysics (INAF), I-00136 Rome, Italy} 
$^{4}$ {ETH Z\"urich, CH-8093 Z\"urich, Switzerland} 
$^{5}$ {Institut de F\'isica d'Altes Energies (IFAE), The Barcelona Institute of Science and Technology (BIST), E-08193 Bellaterra (Barcelona), Spain} 
$^{6}$ {Japanese MAGIC Group: Institute for Cosmic Ray Research (ICRR), The University of Tokyo, Kashiwa, 277-8582 Chiba, Japan} 
$^{7}$ {Technische Universit\"at Dortmund, D-44221 Dortmund, Germany} 
$^{8}$ {Croatian MAGIC Group: University of Zagreb, Faculty of Electrical Engineering and Computing (FER), 10000 Zagreb, Croatia} 
$^{9}$ {IPARCOS Institute and EMFTEL Department, Universidad Complutense de Madrid, E-28040 Madrid, Spain} 
$^{10}$ {Centro Brasileiro de Pesquisas F\'isicas (CBPF), 22290-180 URCA, Rio de Janeiro (RJ), Brazil} 
% ...
$^{41}$ {also at International Center for Relativistic Astrophysics (ICRA), Rome, Italy} 
% ...

\end{document}